# Eigensequences for Multiuser Communication over the Real Adder Channel

R. M. Campello de Souza, H. M. de Oliveira[1]

*Abstract*—**Shape-invariant signals under the Discrete Fourier Transform are investigated, leading to a class of eigenfunctions for the unitary discrete Fourier operator. Such invariant sequences (eigensequences) are suggested as user signatures over the real adder channel (*t*-RAC) and a multiuser communication system over the *t*-RAC is presented.**

## I. Introduction

Fourier transforms may be looked upon as linear operators. One interesting problem in this context is, in the language of operators, to determine the so-called eigenfunctions [1-2]. Let $\mathfrak{I}$ be a vector space equipped with a linear transformation $\mathsf{F}: \mathfrak{I} \rightarrow \mathfrak{I}$, $v \mapsto \mathsf{F}(v)$; eigenfunctions of $\mathsf{F}$ are solutions of $\mathsf{F}(v) = \lambda v$. In the case of the classical Fourier transform, it implies that $\mathsf{F}\{f(t)\} = \lambda . f(w)$, for some $f \in L^2(R)$, $\lambda$ a scalar. This means that the information time and frequency representations share the same shape [3].

This paper investigates eigensequences of the Discrete Fourier Transform (DFT) operator. A sequence whose shape remains unchanged by the DFT operator is said to be a DFT invariant sequence (DFTIS). Such sequences are attractive in applications involving spectral analysis, since the computational complexity of its DFT is linear.

In next section, the unitary DFT operator is presented and new systematic procedures to generate a DFTIS are introduced. It is shown that the set of all DFTISs of length $N$ and associated eigenvalue $\lambda$ is a vector subspace over the field of real numbers. Section 3 proposes a new multiuser communication system over the 2, 3, and 4-user real adder channel (/2/3/4/-RAC). The conclusions are presented in section 4.

## II. Eigensequences of the DFT

In what follows, the unitary form of the DFT operator is considered.

*Definition 1.* (Discrete Fourier Transform [4]) The DFT of $x[n]$, a length $N$ sequence of real or complex components, is the sequence $X[k]$ of elements

$$X[k] = \frac{1}{\sqrt{N}} \sum_{n=0}^{N-1} x[n] W^{kn}, \; k=0,..., N-1, \qquad (1)$$

where $W = e^{-j2\pi/N}$. The inverse DFT of $X[k]$ is

$$x[n] = \frac{1}{\sqrt{N}} \sum_{k=0}^{N-1} X[k] W^{-kn}, \; n=0,..., N-1. \qquad (2)$$

The DFT pair is denoted by $x[n] \leftrightarrow X[k]$. Some useful DFT properties are listed below:

P1. $x[n-m] \leftrightarrow W^{mk} X[k]$.

P2. $W^{-rn} x[n] \leftrightarrow X[k-r]$.

P3. $X[n] \leftrightarrow x[-k]$.

P4. $\mathrm{E}\{x[n]\} \leftrightarrow \mathrm{E}\{X[k]\}$.

P5. $\mathrm{O}\{x[n]\} \leftrightarrow \mathrm{O}\{X[k]\}$,

where $\mathrm{E}\{x[n]\}$ and $\mathrm{O}\{x[n]\}$ denote, respectively, the even and odd parts of $x[n]$. All the sequences considered in this paper have length $N$, which means that the arithmetic in their indexes is modulo $N$.

This paper investigates sequences $x[n]$ whose DFT $X[k]$ has the form

$$X[k] = \lambda x[k]. \qquad (3)$$

In operator form, $\mathsf{D}\{x[n]\} = \lambda x[n]$, where $\mathsf{D}$ denotes the unitary DFT operator. Such sequences, therefore, are eigensequences of the DFT, with associated eigenvalue $\lambda$, and are denoted by $\lambda N$ -DFTIS. The values of $\lambda$ satisfying (3) are given by proposition 1.

*Proposition 1.* The eigenvalues of the unitary DFT operator are the roots of unity of order four ($\pm 1, \pm j$).

*Proof:* Applying four times the DFT operator to $x[n]$ and considering property P3 (duality), leads to the following DFT pairs:

$$x[n] \leftrightarrow X[k],$$
$$X[n] \leftrightarrow x[-k],$$
$$x[-n] \leftrightarrow X[-k],$$
$$X[-n] \leftrightarrow x[k],$$

which means that

$$\mathsf{D}^{(4)}\{x[n]\} = x[k].$$

[1] This work was supported in part by Conselho Nacional de Desenvolvimento Científico e Tecnológico (CNPq) proccess#306180. The authors are with Universidade Federal de Pernambuco, Departamento de Eletrônica e Sistemas, Grupo de Comunicações, C.P. 7800, 50.711-970, Recife – PE. (e-mail: {ricardo,hmo}@ ufpe.br).

Since $x[n]$ is a DFTIS, the above expression is equivalent to $\lambda^4 x[k]=x[k]$ and the result follows.

■

Proposition 2 below sheds some light into the nature of a $\lambda N$-DFTIS.

*Proposition 2.* If $x[n]$ is a $\lambda N$-DFTIS, then $x[n]$ is, respectively, an even or odd sequence according to $\lambda=\pm 1$ or $\lambda=\pm j$.

*Proof:* If $x[n]\leftrightarrow X[k]$ then, considering that $x[n]$ is a $\lambda N$-DFTIS, the resulting pair from P3 is $\lambda x[n]\leftrightarrow x[-k]$. Therefore $\lambda^2 x[k]=x[-k]$, so that if $\lambda=\pm 1$, $x[n]=x[-n]$ and, if $\lambda=\pm j$, $x[n]=-x[-n]$. ■

Odd and even sequences may be used to generate $\lambda N$-DFTISs, according to propositions 3 and 4 below.

*Proposition 3.* If $x[n]$ is a length $N$ sequence, then the sequence $y[n]=\mathrm{E}\{x[n]\}\pm \mathrm{E}\{X[n]\}$ is a $\lambda N$-DFTIS, with associated eigenvalue $\lambda=\pm 1$.

*Proof:* From properties P3 and P4, emerges the pair

$$\mathrm{E}\{x[n]\}\pm \mathrm{E}\{X[n]\}\leftrightarrow \mathrm{E}\{X[k]\}\pm \mathrm{E}\{x[-k]\},$$

i.e.,

$$\mathrm{E}\{x[n]\}\pm \mathrm{E}\{X[n]\}\leftrightarrow \pm\{\mathrm{E}\{x[k]\}\pm \mathrm{E}\{X[k]\}\},$$

and $y[n]$ is a DFTIS with associated eigenvalue $\lambda=\pm 1$.

■

*Corollary:* Every even sequence $x[n]$ yields a DFTIS $y[n]=x[n]\pm X[n]$. ■

*Proposition 4.* If $x[n]$ is a length $N$ sequence, then the sequence $y[n]=\mathrm{O}\{x[n]\}\mp j\mathrm{O}\{X[n]\}$ is a $\lambda N$-DFTIS, with associated eigenvalue $\lambda=\pm j$.

*Proof:* From properties P3 and P5, emerges the pair

$$\mathrm{O}\{x[n]\}\mp j\mathrm{O}\{X[n]\}\leftrightarrow \mathrm{O}\{X[k]\}\mp j\mathrm{O}\{x[-k]\},$$

i.e.,

$$\mathrm{O}\{x[n]\}\mp j\mathrm{O}\{X[n]\}\leftrightarrow \pm j\{\mathrm{O}\{x[k]\}\mp j\mathrm{O}\{X[k]\}\}$$

and $y[n]$ is a DFTIS with associated eigenvalue $\lambda=\pm j$.

■

*Corollary:* Every odd sequence $x[n]$ yields a DFTIS $y[n]=x[n]\mp jX[n]$.

■

Propositions 3 and 4 indicate how a DFTIS may be constructed from an even or odd sequence. From one such sequence it is possible to construct families of $\lambda N$- DFTISs, according to proposition 5.

*Proposition 5.* If $x[n]$ is a $\lambda N$-DFTIS, then the sequence $g[n,m]=x[n+m]+2\cos[\frac{2\pi}{N}mn]x[n]+x[n-m]$ is also a $\lambda N$-DFTIS, for any integer *m*; $g[n,m]$ is said to be a generating function of $\lambda N$-DFTISs.

*Proof:* The DFT of $g[n,m]$ is

$$G[k,m]=W^{-km}X[k]+X[k-m]+X[k+m]+W^{km}X[k].$$

Considering that $X[k]=\lambda x[k]$, leads to

$$G[k,m]=\lambda(x[k+m]+2\cos[\frac{2\pi}{N}mk]x[k]+x[k-m])$$

and $g[n,m]$ is a $\lambda N$-DFTIS. ■

Sets of DFTISs engender well-known algebraic structures with respect to componentwise addition of sequences. The properties of these structures play an important part in the communication system described in section III. Let $I_N^\lambda$ denote the set of all $\lambda N$-DFTISs.

*Proposition 6.*

i) $I_N^\lambda$ is a group with respect to componentwise real addition (denoted $\oplus$) of sequences.

ii) $I_N^\lambda$ is a vector space over the field of real numbers, with respect to $\oplus$, denoted by $V_N^\lambda$.

*Proof:* i) The algebraic structure $< I_N^\lambda, \oplus >$ satisfies the four axioms of a group definition.

ii) Since the DFT operator is linear, if $x_1[n]$ and $x_2[n]\in V_N^\lambda$, then the sequence $x_1[n]\oplus x_2[n]\in V_N^\lambda$. Also, for any real number *a*, the sequence $a\,x_1[n]\in V_N^\lambda$ ■

For each one of the eigenvalues $r=\pm 1,\pm j$, proposition 6 defines a group of $\lambda N$-DFTISs, which are denoted by $G_N^+$, $G_N^-$, $G_N^j$ and $G_N^{-j}$, respectively and also a vector subspace, which are denoted by $V_N^+$, $V_N^-$, $V_N^j$ and $V_N^{-j}$, respectively.

## III. SEQUENCES FOR THE REAL ADDER CHANNEL

The generating function from proposition 5 (reproduced here),

$$g[n,m]=x[n+m]+2\cos(\frac{2\pi}{N}mn)x[n]+x[n-m],$$

is periodic, with period *N*, and satisfies $g[n,m]=g[n,N-m]$. Therefore, for a given sequence $x[n]$, $g[n,m]$ generates $\lceil N/2\rceil$ length $N$ distinct invariant sequences, for $m=1,2,.....,\lceil N/2\rceil$ ($\lceil . \rceil$ denotes the ceiling function). In general, the components of such sequences will not always be integers. However, if $N$ is a perfect square, all integer sequences may be obtained from $g[n,m]$.

### A. *DFT Invariant Sequences*

A list of some invariant sequences for several values of $N$, generated by $g[n,m]$, is given in tables I and II.

TABLE I.
DFT INVARIANT SEQUENCES (NON-INTEGER COMPONENTS)

| $N$ | $\lambda$ | Sequence |
|---|---|---|
| 5 | 1 | 8.4721.. 4.8541.. 0.3820.. 0.3820.. 4.8541.. |
| 5 | -1 | 0.4721.. -0.3820.. -0.3820.. -0.3820.. -0.3820.. |
| 8 | 1 | 9.6569.. 2 2.8284.. 2 4 2 2.8284.. 2 |
| 8 | -1 | 1.6569.. -0.5858.. -2 -0.5858.. 0 -0.5858.. -2 -0.5858.. |
| 9 | 1 | 10 2.3473.. 3.1206.. 1 3.5321.. 3.5321.. 1 3.1206.. 2.3473 |
| 9 | -1 | 2 -0.5321.. -2.3473.. -1 -0.1206.. -0.1206.. -1 -2.3473.. -0.5321.. |
| 16 | 1 | 12 3.4142.. 6 0.5858.. 0 0.5858.. 2 3.4142.. 4 3.4142.. 2 0.5858.. 0 0.5858.. 6 3.4142.. |
| 16 | -1 | 4 -2.7654.. -0.5858.. 3.8478.. -2 -3.8478.. -3.4142.. -1.2346.. 0 -1.2346.. -3.4142.. -3.8478.. -2 3.8478.. -0.5858.. -2.7654.. |

TABLE II.
DFT INVARIANT SEQUENCES (INTEGER COMPONENTS)

| $N$ | $\lambda$ | Sequence |
|---|---|---|
| 4 | 1 | 4 1 2 1 |
| 4 | -1 | 2 -2 -2 -2 |
| 9 | 1 | 10 1 1 7 1 1 7 1 1 |
| 9 | -1 | 2 -1 -1 -1 -1 -1 -1 -1 -1 |
| 16 | 1 | 12 2 0 2 8 2 0 2 4 2 0 2 8 2 0 2 |
| 16 | -1 | 4 -2 0 -2 0 -2 0 -2 -4 -2 0 -2 0 -2 0 -2 |
| 25 | 1 | 6 1 1 1 1 1 1 1 1 1 1 1 1 1 1 1 1 1 1 1 1 1 1 1 1 |
| 25 | -1 | 4-1-1-1-1-1-1-1-1-1-1-1-1-1-1-1-1-1-1-1-1-1-1-1-1 |

### *B. The Real Adder Channel*

A well-known communication channel model is the two-user adder binary channel, or 2-BAC [5-6]. In this paper, the /2/3/4/-RAC model is focused, where addition is over the real numbers. DFTISs like those shown in tables I and II may be used for information transmission of two to four users over the RAC.

In what follows $x_i[n]$ denotes the sequence of user *i*. The associated vector spaces are those defined in proposition 6. The real addition of user sequences is denoted by $y[n]$.

#### *B1. 2-RAC*

Starting with the 2-RAC, let $x_1[n]$ and $x_2[n]$ be invariant sequences in $V_N^+$ and $V_N^-$, associated with users 1 and 2, respectively. From the sequence obtained from the addition $x_1[n]+x_2[n]$ (from now on componentwise addition of sequences is simply denoted by + instead of $\oplus$), it is possible to recover (separate) the user sequences. From the DFT pairs, $x_1[n] \leftrightarrow X_1[k]$, $x_2[n] \leftrightarrow X_2[k]$ and $y[n] \leftrightarrow Y[k]$, we may write

$$y[n] = x_1[n] + x_2[n], \quad (4a)$$

$$Y[k] = X_1[k] + X_2[k]. \quad (4b)$$

Since $x_1[n]$ and $x_2[n]$ are in $V_N^+$ and $V_N^-$, respectively, (4b) is equivalent to

$$Y[k] = x_1[k] - x_2[k]. \quad (5)$$

From expressions (4a) and (5), the user sequences may be recovered from

$$x_1[n] = \frac{y[n]+Y[n]}{2}, \quad (6a)$$

$$x_2[n] = \frac{y[n]-Y[n]}{2}, \quad (6b)$$

where FFT algorithms may be used to compute the $Y[k]$.

#### *B2 3-RAC*

Now there are $C_4^3 = 4$ possibilities, which are denoted by 123, 124, 134 and 234, where $1 \equiv V_N^+$, $2 \equiv V_N^-$, $3 \equiv V_N^j$ and $4 \equiv V_N^{-j}$.

*The case 123*: Applying twice the DFT operator to $y[n]$, leads to

$$x_1[n] + x_2[n] + x_3[n] = y[n],$$
$$x_1[k] - x_2[k] + jx_3[k] = \mathcal{D}\,(y[n]),$$
$$x_1[k] + x_2[k] - x_3[k] = \mathcal{D}^2\,(y[n]).$$

The coefficient matrix is

$$M_3^{123} = \begin{bmatrix} 1 & 1 & 1 \\ 1 & -1 & j \\ 1 & 1 & -1 \end{bmatrix}; \text{ its determinant is } \Delta_3^{123} = 4.$$

The solution of the above system is:

$$x_1[n] = \tfrac{1}{2}\{\mathrm{E}(y[n]) - j\mathrm{O}(y[n]) + Y[n]\},$$
$$x_2[n] = \tfrac{1}{2}\{\mathrm{E}(y[n]) + j\mathrm{O}(y[n]) - Y[n]\},$$
$$x_3[n] = \mathrm{O}\,(y[n]).$$

For the remaining cases, we list the system, the coefficient matrix $M_3$ and the respective solution.

*The case 124*:

$$x_1[n] + x_2[n] + x_3[n] = y[n],$$
$$x_1[k] - x_2[k] - jx_3[k] = \mathcal{D}\,(y[n]),$$
$$x_1[k] + x_2[k] - x_3[k] = \mathcal{D}^2\,(y[n]).$$

$$M_3^{124} = \begin{bmatrix} 1 & 1 & 1 \\ 1 & -1 & -j \\ 1 & 1 & -1 \end{bmatrix}, \ \Delta_3^{124} = 4.$$

System solution:

$$x_1[n] = \tfrac{1}{2}\{\mathrm{E}(y[n]) + j\mathrm{O}(y[n]) + Y[n]\},$$
$$x_2[n] = \tfrac{1}{2}\{\mathrm{E}(y[n]) - j\mathrm{O}(y[n]) - Y[n]\},$$
$$x_3[n] = \mathrm{O}\,(y[n]).$$

*The case 134*:

$$x_1[n] + x_2[n] + x_3[n] = y[n],$$
$$x_1[k] + jx_2[k] - jx_3[k] = \mathcal{D}\,(y[n]),$$
$$x_1[k] - x_2[k] - x_3[k] = \mathcal{D}^2\,(y[n]).$$

$$M_3^{134} = \begin{bmatrix} 1 & 1 & 1 \\ 1 & j & -j \\ 1 & -1 & -1 \end{bmatrix}; \ \Delta_3^{134} = -4j .$$

System solution:

$$x_1[n] = \mathrm{E}(y[n]),$$
$$x_2[n] = \tfrac{1}{2}\{\mathrm{O}(y[n] + j\mathrm{E}(y[n]) - jY[n]\},$$
$$x_3[n] = \tfrac{1}{2}\{\mathrm{O}(y[n] - j\mathrm{E}(y[n]) + j(y[n]\}.$$

*The case 234*:

$$x_1[n] + x_2[n] + x_3[n] = y[n],$$
$$-x_1[k] + jx_2[k] - jx_3[k] = \mathcal{D}\ (y[n]),$$
$$x_1[k] - x_2[k] - x_3[k] = \mathcal{D}^2\ (y[n]).$$

$$M_3^{234} = \begin{bmatrix} 1 & 1 & 1 \\ -1 & j & -j \\ 1 & -1 & -1 \end{bmatrix}; \ \Delta_3^{234} = -4j .$$

System solution:

$$x_1[n] = \mathrm{E}(y[n]),$$
$$x_2[n] = \tfrac{1}{2}\{\mathrm{O}(y[n]) - j\mathrm{E}(y[n]) - jY[n]\},$$
$$x_3[n] = \tfrac{1}{2}\{\mathrm{O}(y[n]) + j\mathrm{E}(y[n]) + jY[n]\}.$$

*B3. 4-RAC*

Applying three times the DFT operator to $y[n]$, leads to the following system:

$$x_1[n] + x_2[n] + x_3[n] + x_4[n] = y[n],$$
$$x_1[k] - x_2[k] + jx_3[k] - jx_3[k] = \mathcal{D}(y[n]),$$
$$x_1[k] + x_2[k] - x_3[k] - x_4[k] = \mathcal{D}^2(y[n]),$$
$$x_1[k] - x_2[k] - jx_3[k] + jx_4[k] = \mathcal{D}^3(y[n]).$$

The coefficient matrix is $M_4 = \begin{bmatrix} 1 & 1 & 1 & 1 \\ 1 & -1 & j & -j \\ 1 & 1 & -1 & -1 \\ 1 & -1 & -j & j \end{bmatrix}; \ \Delta_4 = 16j,$

and the system solution is:

$$x_1[n] = \tfrac{1}{2}\{\mathrm{E}(y[n]) + \mathrm{E}(Y[n])\},$$
$$x_2[n] = \tfrac{1}{2}\{\mathrm{E}(y[n]) - \mathrm{E}(Y[n])\},$$
$$x_3[n] = \tfrac{1}{2}\{\mathrm{O}(y[n]) - j\mathrm{O}(Y[n])\},$$
$$x_4[n] = \tfrac{1}{2}\{\mathrm{O}(y[n]) + j\mathrm{O}(Y[n])\}.$$

*C. Design of a 2-user communication system*

In what follows a 2-user information transmission system, based on the vector spaces $V_N^+$ and $V_N^-$, is proposed. User sequences are weighted and combined in the 2-RAC, resulting in

$$y[n] = a_1 x_1[n] + a_2 x_2[n],$$

where $a_1, a_2$ are real numbers. The system is designed such that $x_1[n] \in V_N^+$ and $x_2[n] \in V_N^-$. In a practical scenario let us suppose that the energy of the channel sequence is bounded by a given value, say $E_{max}$. That implies the existence of limits in the dynamic range of the user coefficients, i.e. $a_1, a_2 \in [-M, +M]$, where $M < +\infty$. A sufficient condition for the choice of the peak value M is established in the following proposition.

*Proposition 7.* Consider a 2-RAC, where the channel sequence energy should not exceed a certain limit value $E_{max}$. If the user sequences $x_1[n] \in V_N^+$ and $x_2[n] \in V_N^-$, are weighted by scalars $a_1, a_2$ with dynamic range $\pm \sqrt{\dfrac{E_{\max}}{\sum_{n=1}^{N}(|x_1[n]| + |x_2[n]|)^2}}$, then the channel sequence energy does not exceed $E_{max}$.

*Proof:* The energy of the channel sequence $y[n]$ is upper bounded by

$$\sum_{n=1}^{N} |y[n]|^2 \le \sum_{n=1}^{N} (|a_1|.|x_1[n]| + |a_2|.|x_2[n]|)^2 .$$

Since $|a_i| \le M$, one may write

$$\sum_{n=1}^{N} |y[n]|^2 \le M^2 \sum_{n=1}^{N} (|x_1[n]| + |x_2[n]|)^2 ,$$

so that, if

$$M^2 \sum_{n=1}^{N} (|x_1[n]| + |x_2[n]|)^2 = E_{\max} ,$$

the result follows.

■

Any two λN-DFTISs may be assigned to the users, as long as $x_1[n] \in V_N^+$ and $x_2[n] \in V_N^-$. For digital transmission of information, the coefficients $a_1, a_2 \in [-M, +M]$ are quantized into $Q = 2^b$ levels (*b* bits A/D converter). The binary sequences of each user are divided into *b* bits words, at each transmission interval *T*, corresponding to the transmission interval of the sequence $x_i[n]$. The user transmission rate is *b* bits/ *T*. Observe that in a noiseless channel, *b* may assume an arbitrarily large value, and the transmission rate is unbounded. A block diagram of a transmission system over the 2-RAC is shown in figure 1. The performance of such systems in noisy conditions has not yet been analyzed. The proposed transmission system can be seen as a generalization of a direct sequence spread spectrum system (DS-CDMA) where, instead of spreading one bit using a signature sequence, *b* bits are spread over a length *N* sequence.

*Example:* To illustrate the ideas discussed above, let us consider the use of λN-DFTISs over a 2-RAC.

a) For $N$=4, with $x_1[n]$=[1 0 1 0] and $x_2[n]$=[1 -1 -1 -1], the user coefficients are bounded by $|a_i| \le \sqrt{E_{\max}/10}$.

b) Suppose that the DFTISs given in (a) are to be used over a 2-RAC with $E_{max}$=10. From proposition 7, $a_1, a_2 \in [-1,+1]$. Considering an A/D converter with $Q$=4 uniform levels, the user coefficients are quantized into $\left\{-1,-\frac{1}{3},\frac{1}{3},+1\right\}$, corresponding to the binary labels {01, 00, 10, 11}. If the sequences to be transmitted by the two users are 0101... and 1110... (fig. 1), the channel sequence transmitted over the 2-RAC will be $y[n]$ = [0,-1,-2,-1 | 2/3,-1/3,-4/3,-1/3 | ...].

### *D.* A *Multiuser Communication System*

A multiuser system can easily be devised. In broad lines, users are divided into 2/3/4 groups, each one assigned to the vector subspaces $V_N^+$, $V_N^-$, $V_N^j$ and $V_N^{-j}$. Each user of a particular group receives a distinct DFTIS, which belongs to the subspace associated with this group, as its signature sequence. A random choice, which may include some priority for users, is then performed, selecting the $t$-users ($t$=2, 3, 4) per channel use, each one belonging to a different group. The selected users get the "*token*" to access the RAC: just $t$ among them access the $t$-RAC per channel intervention, despite the greater number of users in the system. At the receiver, once the signature sequence of each of the $t$ users is recovered, the information about the transmitted message and the user identification associated with it is retrieved.

## IV. CONCLUSIONS

This paper investigates sequences that are invariant to the unitary DFT operator. Such sequences, which were termed DFTIS, are attractive from the point of view that the computation of its DFT requires a linear computational complexity. Systematic procedures for the generation of these sequences were proposed. In particular, a generating function for DFTISs of arbitrary length was introduced. The sets of invariant sequences were characterized in terms of well known algebraic structures.

A multiuser communication system for the Real Adder Channel, for 2, 3 and 4 users was proposed (/2/3/4/-RAC). The systems apply DFTISs as user signatures. In each case, expressions for the recovery of the user information were given (table III). The performance of such schemes is currently under investigation.

## ACKNOWLEDGMENTS

The authors would like to thank Dr. Renato J. S. Cintra for valuable suggestions.

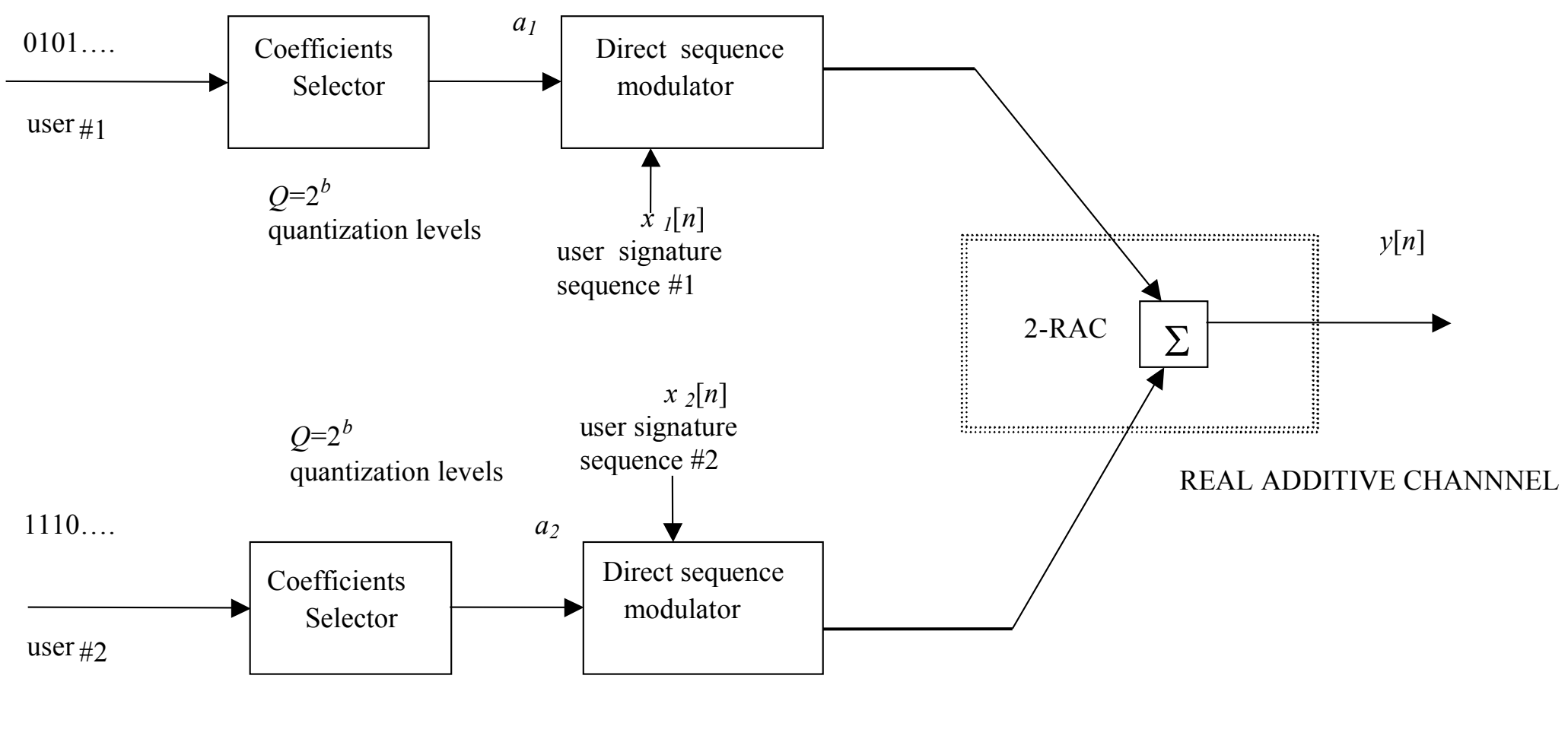

**Fig. 1**. Direct eigensequences transmission system for two users over the 2-RAC. User signature sequences are DFTISs of length $N$, belonging to the subspaces of real eigenvalues $\pm 1$. The coefficient selection is made by a $b$ bits D/A converter.

TABLE III.
USER RECOVERY ON DFTIS BASED MULTIUSER COMMUNICATION SYSTEM (/2/3/4-RAC).

| us/se | user 1 | user 2 | user 3 | user 4 |
|---|---|---|---|---|
| 2 | $x_1[n]=\frac{y[n]+Y[n]}{2}$ | $x_2[n]=\frac{y[n]-Y[n]}{2}$ | - | - |
| 3/123 | $x_1[n]=$ $\frac{1}{2}\{\mathrm{E}(y[n])-j\mathrm{O}(y[n])+Y[n]\}$ | $x_2[n]=$ $\frac{1}{2}\{\mathrm{E}(y[n])+j\mathrm{O}(y[n])-Y[n]\}$ | $x_3[n]=\mathrm{O}\,(y[n])$ | - |
| 3/124 | $x_1[n]=$ $\frac{1}{2}\{\mathrm{E}(y[n])+j\mathrm{O}(y[n])+Y[n]\}$ | $x_2[n]=$ $\frac{1}{2}\{\mathrm{E}(y[n])-j\mathrm{O}(y[n])-Y[n]\}$ | $x_3[n]=\mathrm{O}\,(y[n])$ | - |
| 3/134 | $x_1[n]=\mathrm{E}(y[n])$ | $x_2[n]=$ $\frac{1}{2}\{\mathrm{O}(y[n])+j\mathrm{E}(y[n])-jY[n]\}$ | $x_3[n]=$ $\frac{1}{2}\{\mathrm{O}(y[n])-j\mathrm{E}(y[n])+j(y[n])\}$ | - |
| 3/234 | $x_1[n]=\mathrm{E}(y[n])$ | $x_2[n]=$ $\frac{1}{2}\{\mathrm{O}(y[n])-j\mathrm{E}(y[n])-jY[n]\}$ | $x_3[n]=$ $\frac{1}{2}\{\mathrm{O}(y[n])+j\mathrm{E}(y[n])+jY[n]\}$ | |
| 4 | $x_1[n]=$ $\frac{1}{2}\{\mathrm{E}(y[n])+\mathrm{E}(Y[n])\}$ | $x_2[n]=$ $\frac{1}{2}\{\mathrm{E}(y[n])-\mathrm{E}(Y[n])\}$ | $x_3[n]=$ $\frac{1}{2}\{\mathrm{O}(y[n])-\mathrm{jO}(Y[n])\}$ | $x_4[n]=$ $\frac{1}{2}\{\mathrm{O}(y[n])+\mathrm{jO}(Y[n])\}$ |

us=user; se=index of the vector space.